\documentclass[a4paper,dvips]{ifacconf}

\usepackage{natbib}            
\usepackage{graphicx}          
\usepackage{amsmath,amsfonts,amssymb}

\newcommand{\samp}[1]{{\mathcal S}_{#1}}
\newcommand{\hold}[1]{{\mathcal H}_{#1}}
\newcommand{\E}{{\mathcal E}}
\newcommand{\K}{{\mathcal K}}

\newcommand{\ee}{{\mathrm{e}}}
\newcommand{\jj}{{\mathrm{j}}}

\newcommand{\one}{{\mathbf{1}}}
\newcommand{\abcd}[4]{\left[\begin{array}{c|c}#1&#2\\\hline#3&#4\end{array}\right]}

\def\lift{\mathop{\bf lift}\nolimits}

\def\blkdiag{\mathop{\bf blkdiag}\nolimits}
\newcommand{\mr}{{\mathrm{mr}}}

\newcommand{\dlift}[1]{{\mathbf L}_{#1}}
\newcommand{\idlift}[1]{{\mathbf L}_{#1}^{-1}}
\newcommand{\us}[1]{\uparrow\!#1}
\newcommand{\ds}[1]{\downarrow\!#1}

\newtheorem{problem}{Problem}

\begin{document}

\begin{frontmatter}

\title{FIR Digital Filter Design by Sampled-Data $H^\infty$ Discretization%
\thanksref{footnoteinfo}} 

\thanks[footnoteinfo]{This research is supported in part by the JSPS Grant-in-Aid for Scientific Research (B) No.~24360163
and (C) No.~24560543, and Grant-in-Aid for Exploratory Research No.~22656095.}

\author[MN]{Masaaki Nagahara} 
\author[YY]{Yutaka Yamamoto} 

\address[MN]{
Graduate School of Informatics, 
        Kyoto University,\\
        Sakyo-ku Yoshida-Honmachi, Kyoto 606-8501, JAPAN\\
        (e-mail: nagahara@ieee.org)}
\address[YY]{
Graduate School of Informatics, 
        Kyoto University,\\
        Sakyo-ku Yoshida-Honmachi, Kyoto 606-8501, JAPAN\\
        (e-mail: yy@i.kyoto-u.ac.jp)}
          

\begin{abstract}                          
FIR (finite impulse response) digital filter design is a fundamental problem in signal processing.
In particular, FIR approximation of analog filters (or systems) is
ubiquitous not only in signal processing but also in digital
implementation of controllers.
In this article, we propose a new design method of an FIR digital filter
that optimally approximates a given analog filter
in the sense of minimizing the $H^\infty$ norm
of the sampled-data error system.
By using the lifting technique
and the KYP (Kalman–Yakubovich–Popov) lemma, 
we reduce the $H^\infty$ optimization to a convex optimization described by an LMI (linear matrix inequality).
We also extend the method to multi-rate and multi-delay systems.
A design example is shown to illustrate the effectiveness of the proposed method.
\end{abstract}

\end{frontmatter}

\section{Introduction}
In this article, we consider a fundamental problem in signal processing, namely
FIR (finite impulse response) approximation of \emph{analog} filters.
FIR digital filters are preferred to IIR (infinite-impulse response) digital filters
because of the following merits:
\begin{itemize}
\item FIR filters are always stable.
 \item They can be easily implemented in digital systems.
 \item They are free from problems of IIR filters such as limit cycles caused by quantization.
\end{itemize}
On the other hand, there are  a considerable design methods
for IIR digital filters, e.g., Butterworth, Chebyshev and Elliptic, to name a few
(see \cite{OppSch}).
To obtain an FIR digital filter that approximates a given IIR \emph{digital} filter,
approximation methods with an appropriate optimization 
have been proposed in 
\cite{KooBitGre92,YamAndNagKoy03} for example.
These methods are available if we are given a target IIR digital filter.

In practice, a target filter (or a system)
to be approximated by an FIR digital filter
may be at first given by an \emph{analog} filter.
An RLC filter (an electrical circuit consisting of resistors,
inductors, and capacitors)
is one example and
a PID (proportional-integral-derivative) controller is another.
To obtain an FIR digital filter that mimics such an analog filter,
one might go through two steps: 
\begin{enumerate}
\item Compute an IIR digital filter
that approximates the original analog filter via
step-invariant transformation or bilinear (or Tustin) transformation
(see \cite{CheFra}),
\item Approximate the obtained IIR digital filter to an FIR one
by truncation of the impulse response, or
more sophisticated methods as in \cite{KooBitGre92,YamAndNagKoy03}.
\end{enumerate}

Obviously, it is desirable if an FIR digital filter is obtained \emph{directly} from the original analog filter.
For this purpose, we propose a direct design method of FIR digital filters based on
the theory of sampled-data $H^\infty$ control.
We have proposed a design method in \cite{NagYam13}
via sampled-data $H^\infty$ control theory,
which gives the IIR digital filter that approximates a given analog filter
with the $H^\infty$ performance index.
In this article, we extend this work to FIR digital filter design.
A key idea is to use the KYP (Kalman-Yakubovich-Popov) lemma
that reduces the $H^\infty$ optimization problem to an optimization
described in an LMI (linear matrix inequality).
We also extend the result to multi-rate systems that consists
an up-sampler and a fast hold,
and to multi-delay systems that appear in
the Smith predictor proposed in \cite{Smi57},
or multipath propagation in wireless communications (see \cite{Gol} for example).
A design example is shown to illustrate the effectiveness of our methods.

\section{Mathematical Notation and Review}
Throughout this article, we use the following notation.
$L^2[0,\infty)$ is the Lebesgue space consisting of all square integrable real functions on $[0,\infty)$.
$L^2[0,\infty)$ is sometimes abbreviated to $L^2$.
The $L^2$ norm of $f\in L^2$ is defined by
\[
 \|f\|_2 := \sqrt{\int_0^\infty |f(t)|^2dt}.
\]
The symbol $t$ denotes the argument of time,
$s$ the argument of Laplace transform,
and $z$ the argument of $Z$ transform.
These symbols are used to indicate whether a signal or a system is of continuous-time or discrete-time;
for example, $y(t)$ is a continuous-time signal,
$F(s)$ is a continuous-time system,
$K(z)$ is a discrete-time system.
The operator $\ee^{-ls}$ with nonnegative integer $l$ denotes continuous-time delay (or shift) operator:
$(\ee^{-ls}y)(t)=y(t-l)$.
$\samp{h}$  and $\hold{h}$ denote the ideal sampler and the zero-order hold
respectively with sampling period $h>0$.
We denote the imaginary number $\sqrt{-1}$ by $\jj$.

A transfer function with state-space matrices $A,B,C,D$ is denoted by
\[
 \abcd{A}{B}{C}{D} 
  := \begin{cases}
    C(sI-A)^{-1}B+D,&\text{~(continuous-time)}\\
    C(zI-A)^{-1}B+D,&\text{~(discrete-time)}
  \end{cases}
\]
For this notation, we have the following formulae:
\begin{equation}
 \abcd{A_1}{B_1}{C_1}{D_1} \pm \abcd{A_2}{B_2}{C_2}{D_2}
  = \left[\begin{array}{cc|c}A_1&0&B_1\\0&A_2&\pm B_2\\\hline C_1&C_2&D_1\pm D_2\end{array}\right],
  \label{eq:ss_pm}
\end{equation}
\begin{equation}
 \abcd{A_1}{B_1}{C_1}{D_1} \times \abcd{A_2}{B_2}{C_2}{D_2}
  = \left[\begin{array}{cc|c}A_2&0&B_2\\B_1C_2&A_1&B_1D_2\\\hline D_1C_2&C_1&D_1D_2\end{array}\right].
  \label{eq:ss_mult}
\end{equation}
For a stable discrete-time system $G(z)$,
its discrete-time $H^\infty$ norm is defined by
\[
 \|G\|_\infty := \max_{\theta \in [0,\pi]} \sigma_{\max}(G(e^{j\theta})),
\]
where $\sigma_{\max}(\cdot)$ is the largest singular value of the argument matrix.
The well-known  KYP (Kalman-Yakubovich-Popov) lemma
(see \cite{And67,Ran96,TuqVai98,Nag11})
characterizes
the $H^\infty$ norm of a discrete-time system by
a linear matrix inequality (LMI):
\begin{lem}[KYP lemma]
\label{lem:KYP}
Let ${A,B,C,D}$ be a minimal realization of a 
stable discrete-time transfer function $G(z)$.
Let $\gamma>0$.
Then the following are equivalent conditions:
\begin{enumerate}
 \item $\|G\|_\infty = \left\|\abcd{A}{B}{C}{D}\right\|_\infty< \gamma$.
 \item There exists a positive definite matrix $X$ such that
\[
  \left[\begin{array}{ccc}
   A^\top XA - X & A^\top XB & C^\top \\
   B^\top XA & B^\top XB-\gamma I & D^\top\\
   C & D & -\gamma I
   \end{array}\right]
  <0.
\]
 \end{enumerate}
\end{lem}
This lemma is a key to derive
a computationally efficient method
for FIR digital filter design.

\section{Problem Formulation}
\label{sec:problem}
\begin{figure}[tb]
 \centering
\unitlength 0.1in
\begin{picture}( 28.0000,  4.0000)(  4.0000, -6.0000)
%
\special{pn 8}%
\special{pa 400 400}%
\special{pa 800 400}%
\special{fp}%
\special{sh 1}%
\special{pa 800 400}%
\special{pa 734 380}%
\special{pa 748 400}%
\special{pa 734 420}%
\special{pa 800 400}%
\special{fp}%
%
\special{pn 8}%
\special{pa 800 200}%
\special{pa 1200 200}%
\special{pa 1200 600}%
\special{pa 800 600}%
\special{pa 800 200}%
\special{pa 1200 200}%
\special{fp}%
%
\special{pn 8}%
\special{pa 1200 400}%
\special{pa 1600 400}%
\special{dt 0.045}%
\special{sh 1}%
\special{pa 1600 400}%
\special{pa 1534 380}%
\special{pa 1548 400}%
\special{pa 1534 420}%
\special{pa 1600 400}%
\special{fp}%
%
\special{pn 8}%
\special{pa 1600 200}%
\special{pa 2000 200}%
\special{pa 2000 600}%
\special{pa 1600 600}%
\special{pa 1600 200}%
\special{pa 2000 200}%
\special{fp}%
%
\special{pn 8}%
\special{pa 2000 400}%
\special{pa 2400 400}%
\special{dt 0.045}%
\special{sh 1}%
\special{pa 2400 400}%
\special{pa 2334 380}%
\special{pa 2348 400}%
\special{pa 2334 420}%
\special{pa 2400 400}%
\special{fp}%
%
\special{pn 8}%
\special{pa 2400 200}%
\special{pa 2800 200}%
\special{pa 2800 600}%
\special{pa 2400 600}%
\special{pa 2400 200}%
\special{pa 2800 200}%
\special{fp}%
%
\special{pn 8}%
\special{pa 2800 400}%
\special{pa 3200 400}%
\special{fp}%
\special{sh 1}%
\special{pa 3200 400}%
\special{pa 3134 380}%
\special{pa 3148 400}%
\special{pa 3134 420}%
\special{pa 3200 400}%
\special{fp}%
\put(10.0000,-4.0000){\makebox(0,0){$\samp{h}$}}%
\put(18.0000,-4.0000){\makebox(0,0){$K(z)$}}%
\put(26.0000,-4.0000){\makebox(0,0){$\hold{h}$}}%
\put(4.0000,-3.5000){\makebox(0,0)[lb]{$u$}}%
\put(13.0000,-3.5000){\makebox(0,0)[lb]{$v$}}%
\put(21.0000,-3.5000){\makebox(0,0)[lb]{$\psi$}}%
\put(30.5000,-3.5000){\makebox(0,0)[lb]{$\hat{y}$}}%
\end{picture}%
 \caption{Digital system $\K$ consisting of ideal sampler $\samp{h}$,
 digital filter $K(z)$, and zero-order hold $\hold{h}$ with sampling period $h$.}
 \label{fig:digital-system}
\end{figure}
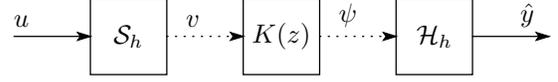

Assume that a transfer function $K_{\rm c}(s)$ of an analog filter is given.
We suppose that $K_{\rm c}(s)$ is a stable%
\footnote{If $K_{\rm c}(s)$ has no poles on the imaginary axis,
the proposed method can be used for unstable $K_{\rm c}(s)$ as follows:
factorize it as $K_{\rm c}(s)=K_{\rm s}(s)K_{\rm as}(s)$,
where $K_{\rm s}(s)$ is stable and $K_{\rm as}(s)$ is
anti-stable,
and then discretize $K_{\rm s}(s)$ and $K_{\rm as}(-s)$
by the proposed method.},
real-rational, proper transfer function.
Our objective is to implement this analog system in a digital system.
To do this, let us consider a digital system shown in Fig.~\ref{fig:digital-system},
where $K(z)$ is an FIR digital filter of length $M$ described by
\begin{equation}
 K(z) = \sum_{k=0}^{M-1} a_kz^{-k},
 \label{eq:FIR}
\end{equation}
$\samp{h}$ is an ideal sampler with a fixed sampling period $h>0$ that
converts a continuous-time signal $u(t)$ 
to a discrete-time signal $v[n]$ as
\[
 v[n] = (\samp{h}u)[n] = u(nh),\quad n=0,1,2,\dots,
\]
and $\hold{h}$ is a zero-order hold
that produces a continuous-time signal $\hat{y}(t)$ 
from a discrete-time signal $\psi[n]$ as
\[
 \hat{y}(t) = \sum_{n=0}^\infty \psi[n]\phi(t-nh),\quad t\in[0,\infty),
\]
where $\phi(t)$ is a box function defined by
\[
 \phi(t) = \begin{cases} 1, & \text{~if~} t\in [0,h),\\ 0, & \text{~otherwise.}\end{cases}
\]

Our problem is to obtain the FIR digital filter coefficients $a_0, a_1, \ldots, a_{M-1}$
in \eqref{eq:FIR} so that the digital system
\[
 \K := \hold{h}K\samp{h} = \hold{h}\left(\sum_{k=0}^{M-1} a_kz^{-k}\right)\samp{h}
\]
mimics the input/output behavior of the analog filter $K_{\rm c}(s)$.

Let $\alpha(t)$ denote the impulse response (or the inverse Laplace transform) of $K_{\rm c}(s)$.
The digital filter $K(z)$ (or the filter coefficients $a_0,a_1,\dots,a_{M-1}$)
is designed to produce a continuous-time signal $\hat{y}$ after
the zero-order hold $\hold{h}$ that
approximates the \emph{delayed} output 
\[
 y(t)=(\alpha\ast u)(t-l)=\int_0^{t-l}\alpha(\tau)u(t-l-\tau)d\tau
\]
of $K_{\rm c}(s)$ for an input $u$.
A positive delay time $l$ may improve the approximation performance when $l$ is large enough
as discussed in e.g., \cite{NagOguYam11,YamNagKha12}.
We assume that $l$ is an integer multiple of $h$, that is, $l=mh$,
where $m$ is a nonnegative integer.
Then, to avoid a trivial solution (i.e., $K(z)=0$),
we should assume some \emph{a priori} information for the input signal $u$.
As used in \cite{NagOguYam11,YamNagKha12},
we adopt the following signal subspace of $L^2[0,\infty)$ to which the input signals belong:
\[
 FL^2 := \left\{Fw: w\in L^2[0,\infty)\right\},
\]
where $F$ is a linear system with a stable, real-rational, strictly proper transfer function $F(s)$.
This transfer function, $F(s)$, defines the \emph{analog characteristic} of the input signals
in the frequency domain.

In summary, our discretization problem is formulated as follows:
\begin{problem}
\label{prob:hinf}
Given target filter $K_{\rm c}(s)$, analog characteristic $F(s)$,
sampling period $h$, and delay step $m$,
find the filter coefficients $a_0,a_1,\ldots,a_{M-1}$ of $K(z)$ 
given by \eqref{eq:FIR} that minimizes
\[
 \begin{split}
 \|\E(K)\|_\infty
  &=\left\|\left(\ee^{-mhs}K_{\rm c}-\hold{h}K\samp{h}\right)F\right\|_\infty\\
  &=\sup_{\substack{w\in L^2\\ w\neq 0}} \frac{\left\|\left(\ee^{-mhs}K_{\rm c}-\hold{h}K\samp{h}\right)Fw\right\|_2}{\|w\|_2}.
 \end{split} 
\]
The corresponding block diagram of the error system
\begin{equation}
 \E(K) := \left(\ee^{-mhs}K_{\rm c}-\hold{h}K\samp{h}\right)F
 \label{eq:error-system}
\end{equation}
is shown in Fig.~\ref{fig:error-system}. 
\begin{figure}[tb]
 \centering
\unitlength 0.1in
\begin{picture}( 32.5000, 10.0000)(  3.0000,-12.0000)
%
\special{pn 8}%
\special{pa 1200 1008}%
\special{pa 1400 1008}%
\special{fp}%
\special{sh 1}%
\special{pa 1400 1008}%
\special{pa 1334 988}%
\special{pa 1348 1008}%
\special{pa 1334 1028}%
\special{pa 1400 1008}%
\special{fp}%
%
\special{pn 8}%
\special{pa 1400 800}%
\special{pa 1800 800}%
\special{pa 1800 1200}%
\special{pa 1400 1200}%
\special{pa 1400 800}%
\special{pa 1800 800}%
\special{fp}%
%
\special{pn 8}%
\special{pa 1800 1000}%
\special{pa 2000 1000}%
\special{dt 0.045}%
\special{sh 1}%
\special{pa 2000 1000}%
\special{pa 1934 980}%
\special{pa 1948 1000}%
\special{pa 1934 1020}%
\special{pa 2000 1000}%
\special{fp}%
%
\special{pn 8}%
\special{pa 2000 800}%
\special{pa 2400 800}%
\special{pa 2400 1200}%
\special{pa 2000 1200}%
\special{pa 2000 800}%
\special{pa 2400 800}%
\special{fp}%
%
\special{pn 8}%
\special{pa 2400 1000}%
\special{pa 2600 1000}%
\special{dt 0.045}%
\special{sh 1}%
\special{pa 2600 1000}%
\special{pa 2534 980}%
\special{pa 2548 1000}%
\special{pa 2534 1020}%
\special{pa 2600 1000}%
\special{fp}%
%
\special{pn 8}%
\special{pa 2600 800}%
\special{pa 3000 800}%
\special{pa 3000 1200}%
\special{pa 2600 1200}%
\special{pa 2600 800}%
\special{pa 3000 800}%
\special{fp}%
\put(16.0000,-10.1000){\makebox(0,0){$\samp{h}$}}%
\put(22.0000,-10.0000){\makebox(0,0){$K(z)$}}%
\put(28.0000,-10.0000){\makebox(0,0){$\hold{h}$}}%
\put(12.7000,-7.0000){\makebox(0,0){$u$}}%
\put(31.0000,-10.5000){\makebox(0,0)[lt]{$\hat{y}$}}%
%
\special{pn 8}%
\special{pa 1200 1008}%
\special{pa 1200 408}%
\special{fp}%
%
\special{pn 8}%
\special{pa 1700 200}%
\special{pa 2100 200}%
\special{pa 2100 600}%
\special{pa 1700 600}%
\special{pa 1700 200}%
\special{pa 2100 200}%
\special{fp}%
%
\special{pn 8}%
\special{pa 2300 200}%
\special{pa 2700 200}%
\special{pa 2700 600}%
\special{pa 2300 600}%
\special{pa 2300 200}%
\special{pa 2700 200}%
\special{fp}%
%
\special{pn 8}%
\special{pa 1200 408}%
\special{pa 1700 408}%
\special{fp}%
\special{sh 1}%
\special{pa 1700 408}%
\special{pa 1634 388}%
\special{pa 1648 408}%
\special{pa 1634 428}%
\special{pa 1700 408}%
\special{fp}%
%
\special{pn 8}%
\special{pa 2100 400}%
\special{pa 2300 400}%
\special{fp}%
\special{sh 1}%
\special{pa 2300 400}%
\special{pa 2234 380}%
\special{pa 2248 400}%
\special{pa 2234 420}%
\special{pa 2300 400}%
\special{fp}%
%
\special{pn 8}%
\special{pa 2700 400}%
\special{pa 3200 400}%
\special{fp}%
%
\special{pn 8}%
\special{pa 1200 700}%
\special{pa 1000 700}%
\special{fp}%
%
\special{pn 8}%
\special{ar 3200 700 50 50  0.0000000  6.2831853}%
%
\special{pn 8}%
\special{pa 3200 400}%
\special{pa 3200 650}%
\special{fp}%
\special{sh 1}%
\special{pa 3200 650}%
\special{pa 3220 584}%
\special{pa 3200 598}%
\special{pa 3180 584}%
\special{pa 3200 650}%
\special{fp}%
%
\special{pn 8}%
\special{pa 3000 1000}%
\special{pa 3200 1000}%
\special{fp}%
%
\special{pn 8}%
\special{pa 3200 1000}%
\special{pa 3200 750}%
\special{fp}%
\special{sh 1}%
\special{pa 3200 750}%
\special{pa 3180 818}%
\special{pa 3200 804}%
\special{pa 3220 818}%
\special{pa 3200 750}%
\special{fp}%
%
\special{pn 8}%
\special{pa 3250 700}%
\special{pa 3550 700}%
\special{fp}%
\special{sh 1}%
\special{pa 3550 700}%
\special{pa 3484 680}%
\special{pa 3498 700}%
\special{pa 3484 720}%
\special{pa 3550 700}%
\special{fp}%
\put(31.0000,-3.5000){\makebox(0,0)[lb]{$y$}}%
\put(34.5000,-6.5000){\makebox(0,0)[lb]{$e$}}%
%
\special{pn 8}%
\special{pa 1000 500}%
\special{pa 600 500}%
\special{pa 600 900}%
\special{pa 1000 900}%
\special{pa 1000 500}%
\special{pa 600 500}%
\special{fp}%
%
\special{pn 8}%
\special{pa 300 700}%
\special{pa 600 700}%
\special{fp}%
\special{sh 1}%
\special{pa 600 700}%
\special{pa 534 680}%
\special{pa 548 700}%
\special{pa 534 720}%
\special{pa 600 700}%
\special{fp}%
\put(8.0000,-7.0000){\makebox(0,0){$F(s)$}}%
\put(19.0000,-4.0000){\makebox(0,0){$K_{\rm c}(s)$}}%
\put(25.0000,-4.0000){\makebox(0,0){$e^{-mhs}$}}%
\put(3.0000,-6.5000){\makebox(0,0)[lb]{$w$}}%
\put(32.5000,-6.5000){\makebox(0,0)[lb]{$+$}}%
\put(32.5000,-7.5000){\makebox(0,0)[lt]{$-$}}%
\end{picture}%

 \caption{Error system $\E(K)$.}
 \label{fig:error-system}
\end{figure}
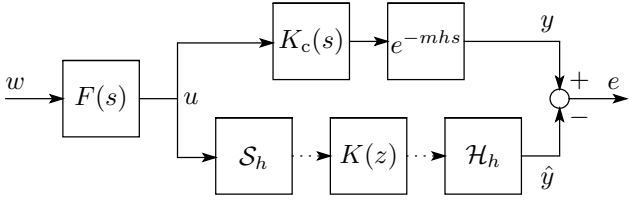
\end{problem}

\section{FIR Filter Design via Sampled-Data $H^\infty$ Optimization}
\label{sec:H-inf}
In this section, we give a design formula to numerically compute the $H^\infty$-optimal filter
coefficients of Problem \ref{prob:hinf} via \emph{fast sample/hold} approximation
as used in \cite{NagYam13},
and the KYP lemma described in Lemma \ref{lem:KYP}.

We first define the discrete-time lifting of a discrete-time system by
\[
 \begin{split}
  &\lift\left(\abcd{A}{B}{C}{D},N\right)\\
  &\quad := \dlift{N}\abcd{A}{B}{C}{D}\idlift{N}\\
  &\quad = \left[\begin{array}{c|cccc}
	 A^N & A^{N-1}B & A^{N-2}B & \ldots & B\\\hline
	 C & D & 0 & \ldots & 0\\
	 CA & CB & D & \ddots & \vdots\\
	 \vdots & \vdots & \vdots & \ddots & 0\\
	 CA^{N-1} & CA^{N-2}B& CA^{N-3}B & \ldots & D
    \end{array}\right],
 \end{split}    
\]
where
\[
\begin{split}
    \dlift{N} &:= (\ds{N})
	\left[\begin{array}{cccc}
	  1 & z & \cdots & z^{N-1}
	\end{array}\right]^T,\\
    \idlift{N} &:=
 	\left[\begin{array}{cccc}
	  1 & z^{-1} & \cdots & z^{-N+1}
	\end{array}\right](\us{N}).
\end{split}
\]
In this definition,
$\ds{N}$ and $\us{N}$ are respectively
a downsampler and an upsampler
(see e.g., \cite{Vai}) defined as
\[
 \begin{split}
  \us{N} &: \bigl\{x[k]\bigr\}_{k=0}^\infty \mapsto \bigl\{x[0],\underbrace{0,\dots,0}_{N-1},x[1],0,\dots\bigr\},\\
  \ds{N} &: \bigl\{x[k]\bigr\}_{k=0}^\infty \mapsto \bigl\{x[0],x[N],x[2N],\dots\bigr\}.
 \end{split}
\]

These operators give the fast sample/hold approximation $E_N$
of the sampled-data error system $\E(K)$ given in \eqref{eq:error-system}
as
\begin{equation}
 E_N(z) = \bigl(z^{-mN}K_{N}(z) - H_NK(z)S_N\bigr)F_N(z),
 \label{eq:EN}
\end{equation}
where
\begin{equation}
 \begin{split}
  K_{N}(z) &= \lift\bigl(\samp{h/N}K_{\rm c}\hold{h/N},N),\\
  F_N(z) &= \lift\bigl(\samp{h/N}F\hold{h/N},N\bigr),\\
  H_N &= [\underbrace{1,\dots,1}_{N}]^\top,\quad
  S_N = [1,\underbrace{0,\dots,0}_{N-1}].
 \end{split}
 \label{eq:formula}
\end{equation}
Note that $K_{N}(z)$ and $F_N(z)$ are finite-dimensional
linear time-invariant systems as shown in \cite{NagYam13}.
Define
\begin{equation}
 \begin{split}
  A_K &:= \left[\begin{array}{ccccc}
  	0&1&0&\ldots&0\\ 
	\vdots&\ddots & \ddots & &\vdots\\
	\vdots& &\ddots&\ddots&0\\
	\vdots& & &\ddots&1\\
	0&\ldots&\ldots&\ldots&0
  \end{array}\right],\quad
  B_K := \left[\begin{array}{c}0\\\vdots\\0\\1\end{array}\right],\\
  C_K &:= \left[\begin{array}{cccc}a_{M-1}&a_{M-2}&\ldots&a_1\end{array}\right],\quad
  D_K := a_0
 \end{split}
 \label{eq:FIR_abcd}
\end{equation}
Then $\{A_K,B_K,C_K,D_K\}$ is a minimal realization of FIR filter $K(z)$ in \eqref{eq:FIR},
that is,
\begin{equation}
 K(z) = \sum_{k=0}^{M-1}a_kz^{-k}=\abcd{A_K}{B_K}{C_K}{D_K}.
 \label{eq:FIR_ss}
\end{equation}
Substituting \eqref{eq:FIR_ss} into \eqref{eq:EN},
we obtain
\[
 E_N(z) = T_1(z) + Q(z) T_2(z),
\]
where
\begin{align}
  T_1(z) &= z^{-mN}K_{N}(z)F_N(z) =: \abcd{A_1}{B_1}{C_1}{0},\label{eq:T1}\\
  T_2(z) &= S_NF_N(z) =: \abcd{A_2}{B_2}{C_2}{0},\label{eq:T2}\\
  Q(z) &= -H_N\sum_{k=0}^{M-1} a_kz^{-k} = \abcd{A_K}{B_K}{H_NC_K}{H_ND_K}.\nonumber
\end{align}
Then using \eqref{eq:ss_pm} and \eqref{eq:ss_mult}, we have
\begin{equation}
 \begin{split}
 E_N(z) &= \left[\begin{array}{ccc|c}
 	A_1 & 0 & 0 & B_1\\
	0 & A_2 & 0 & B_2\\
	0 & B_KC_2 & A_K & 0\\\hline
	C_1 & H_ND_KC_2 & H_NC_K & 0
  \end{array}\right]\\
  &=:\abcd{A}{B}{C(a)}{D(a)},
 \end{split}
 \label{eq:EN_ss}
\end{equation}
where $a$ is the coefficient vector defined by
\begin{equation}
 a = \left[\begin{array}{c}a_0 \\ a_1 \\ \vdots \\ a_{M-1}\end{array}\right] \in {\mathbb R}^M.
 \label{eq:a}
\end{equation}
Note that the coefficient vector $a$ to be designed is independent
of $A$ and $B$ of $E_N(z)$ as in \eqref{eq:EN_ss}.
This property is fundamental for the derivation of LMI optimization.
Finally, our problem (Problem \ref{prob:hinf}) is reduced to
finding $a\in{\mathbb R}^M$ that minimizes the $H^\infty$ norm of $E_N(z)$,
which can be described as an LMI
from the KYP lemma (Lemma \ref{lem:KYP}) as follows.
\begin{problem}
\label{prob:hinf-LMI}
Find $a\in{\mathbb R}^M$ that minimizes $\gamma$ subject to
 \begin{equation}
 \begin{split}
  \left[\begin{array}{ccc}
   A^\top XA - X & A^\top XB & C(a)^\top \\
   B^\top XA & B^\top XB-\gamma I & D(a)^\top\\
   C(a) & D(a) & -\gamma I
   \end{array}\right]
  &<0,\\
  X&>0.
  \end{split}
  \label{eq:hinf_KYP_LMI}
 \end{equation}
\end{problem}
This problem can be efficiently solved via numerical optimization software
such as \texttt{SDPT3} (see \cite{TohTodTut02},
\texttt{SeDuMi} (see \cite{Stu99}),
or \texttt{cvx} (see \cite{cvx,GraBoy08}) on \texttt{MATLAB}.

In summary, we have derived a computationally efficient method with an LMI
in \eqref{eq:hinf_KYP_LMI}
for FIR approximation of analog filters based on fast sample/hold
approximation in \eqref{eq:EN} and the KYP lemma (Lemma \ref{lem:KYP}).

\section{Extension to Multi-rate Systems}
\label{sec:multi-rate}
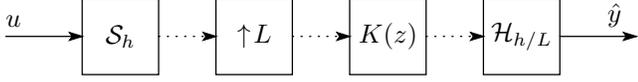
\begin{figure}[tb]
 \centering
\unitlength 0.1in
\begin{picture}( 33.0000,  4.0000)(  4.0000, -6.0000)
%
\special{pn 8}%
\special{pa 400 400}%
\special{pa 800 400}%
\special{fp}%
\special{sh 1}%
\special{pa 800 400}%
\special{pa 734 380}%
\special{pa 748 400}%
\special{pa 734 420}%
\special{pa 800 400}%
\special{fp}%
%
\special{pn 8}%
\special{pa 800 200}%
\special{pa 1200 200}%
\special{pa 1200 600}%
\special{pa 800 600}%
\special{pa 800 200}%
\special{pa 1200 200}%
\special{fp}%
%
\special{pn 8}%
\special{pa 1200 400}%
\special{pa 1500 400}%
\special{dt 0.045}%
\special{sh 1}%
\special{pa 1500 400}%
\special{pa 1434 380}%
\special{pa 1448 400}%
\special{pa 1434 420}%
\special{pa 1500 400}%
\special{fp}%
%
\special{pn 8}%
\special{pa 1500 200}%
\special{pa 1900 200}%
\special{pa 1900 600}%
\special{pa 1500 600}%
\special{pa 1500 200}%
\special{pa 1900 200}%
\special{fp}%
%
\special{pn 8}%
\special{pa 1900 400}%
\special{pa 2200 400}%
\special{dt 0.045}%
\special{sh 1}%
\special{pa 2200 400}%
\special{pa 2134 380}%
\special{pa 2148 400}%
\special{pa 2134 420}%
\special{pa 2200 400}%
\special{fp}%
%
\special{pn 8}%
\special{pa 2200 200}%
\special{pa 2600 200}%
\special{pa 2600 600}%
\special{pa 2200 600}%
\special{pa 2200 200}%
\special{pa 2600 200}%
\special{fp}%
%
\special{pn 8}%
\special{pa 2600 400}%
\special{pa 2900 400}%
\special{dt 0.045}%
\special{sh 1}%
\special{pa 2900 400}%
\special{pa 2834 380}%
\special{pa 2848 400}%
\special{pa 2834 420}%
\special{pa 2900 400}%
\special{fp}%
%
\special{pn 8}%
\special{pa 2900 200}%
\special{pa 3300 200}%
\special{pa 3300 600}%
\special{pa 2900 600}%
\special{pa 2900 200}%
\special{pa 3300 200}%
\special{fp}%
%
\special{pn 8}%
\special{pa 3300 400}%
\special{pa 3700 400}%
\special{fp}%
\special{sh 1}%
\special{pa 3700 400}%
\special{pa 3634 380}%
\special{pa 3648 400}%
\special{pa 3634 420}%
\special{pa 3700 400}%
\special{fp}%
\put(17.0000,-4.0000){\makebox(0,0){$\us{L}$}}%
\put(31.0000,-4.0000){\makebox(0,0){$\hold{h/L}$}}%
\put(24.0000,-4.0000){\makebox(0,0){$K(z)$}}%
\put(10.0000,-4.0000){\makebox(0,0){$\samp{h}$}}%
\put(4.0000,-3.5000){\makebox(0,0)[lb]{$u$}}%
\put(35.5000,-3.5000){\makebox(0,0)[lb]{$\hat{y}$}}%
\end{picture}%
 \caption{Multi-rate system $\K_\mr$ consisting of ideal sampler $\samp{h}$,
 upsampler $\us{L}$, 
 digital filter $K(z)$, and fast hold $\hold{h/L}$.}
 \label{fig:multi-rate-system}
\end{figure}

In this section, we extend the result in the previous section
to multi-rate systems as proposed in \cite{NagYam13}.
That is, we consider the following
multi-rate signal processing system
\[
 \K_\mr := \hold{h/L}K(z)(\us{L})\samp{h}=\hold{h/L}\left(\sum_{k=0}^{M-1}a_kz^{-k}\right)(\us{L})\samp{h},
\]
where $L$ is an integer greater than or equal to $2$.
The block diagram of multi-rate $\K_\mr$ is
shown in Fig.~\ref{fig:multi-rate-system}.
The objective here is to find an FIR digital filter $K(z)$
given by \eqref{eq:FIR}
that minimizes the $H^\infty$ norm of the multi-rate
error system $\E_\mr(K)$ defined by
\begin{equation}
 \E_\mr(K) := \bigl(e^{-mhs}K_{\rm c} - \hold{h/L}K(\us{L})\samp{h}\bigr)F.
 \label{eq:EmrK}
\end{equation}
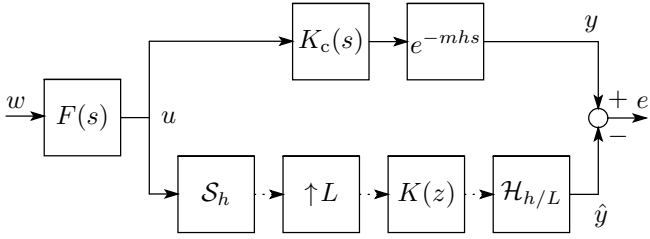
\begin{figure}[tb]
 \centering
\unitlength 0.1in
\begin{picture}( 33.5000, 12.0000)(  3.0000,-14.0000)
%
\special{pn 8}%
\special{pa 1050 1198}%
\special{pa 1200 1198}%
\special{fp}%
\special{sh 1}%
\special{pa 1200 1198}%
\special{pa 1134 1178}%
\special{pa 1148 1198}%
\special{pa 1134 1218}%
\special{pa 1200 1198}%
\special{fp}%
%
\special{pn 8}%
\special{pa 1200 1000}%
\special{pa 1600 1000}%
\special{pa 1600 1400}%
\special{pa 1200 1400}%
\special{pa 1200 1000}%
\special{pa 1600 1000}%
\special{fp}%
%
\special{pn 8}%
\special{pa 1600 1200}%
\special{pa 1750 1200}%
\special{dt 0.045}%
\special{sh 1}%
\special{pa 1750 1200}%
\special{pa 1684 1180}%
\special{pa 1698 1200}%
\special{pa 1684 1220}%
\special{pa 1750 1200}%
\special{fp}%
%
\special{pn 8}%
\special{pa 1750 1000}%
\special{pa 2150 1000}%
\special{pa 2150 1400}%
\special{pa 1750 1400}%
\special{pa 1750 1000}%
\special{pa 2150 1000}%
\special{fp}%
%
\special{pn 8}%
\special{pa 2150 1200}%
\special{pa 2300 1200}%
\special{dt 0.045}%
\special{sh 1}%
\special{pa 2300 1200}%
\special{pa 2234 1180}%
\special{pa 2248 1200}%
\special{pa 2234 1220}%
\special{pa 2300 1200}%
\special{fp}%
%
\special{pn 8}%
\special{pa 2300 1000}%
\special{pa 2700 1000}%
\special{pa 2700 1400}%
\special{pa 2300 1400}%
\special{pa 2300 1000}%
\special{pa 2700 1000}%
\special{fp}%
%
\special{pn 8}%
\special{pa 2700 1200}%
\special{pa 2850 1200}%
\special{dt 0.045}%
\special{sh 1}%
\special{pa 2850 1200}%
\special{pa 2784 1180}%
\special{pa 2798 1200}%
\special{pa 2784 1220}%
\special{pa 2850 1200}%
\special{fp}%
%
\special{pn 8}%
\special{pa 2850 1000}%
\special{pa 3250 1000}%
\special{pa 3250 1400}%
\special{pa 2850 1400}%
\special{pa 2850 1000}%
\special{pa 3250 1000}%
\special{fp}%
\put(19.5000,-12.0000){\makebox(0,0){$\us{L}$}}%
\put(30.5000,-12.0000){\makebox(0,0){$\hold{h/L}$}}%
\put(25.0000,-12.0000){\makebox(0,0){$K(z)$}}%
\put(14.0000,-12.0000){\makebox(0,0){$\samp{h}$}}%
%
\special{pn 8}%
\special{pa 1050 1198}%
\special{pa 1050 398}%
\special{fp}%
%
\special{pn 8}%
\special{pa 1050 398}%
\special{pa 1800 398}%
\special{fp}%
\special{sh 1}%
\special{pa 1800 398}%
\special{pa 1734 378}%
\special{pa 1748 398}%
\special{pa 1734 418}%
\special{pa 1800 398}%
\special{fp}%
%
\special{pn 8}%
\special{pa 1800 200}%
\special{pa 2200 200}%
\special{pa 2200 600}%
\special{pa 1800 600}%
\special{pa 1800 200}%
\special{pa 2200 200}%
\special{fp}%
%
\special{pn 8}%
\special{pa 2200 400}%
\special{pa 2400 400}%
\special{fp}%
\special{sh 1}%
\special{pa 2400 400}%
\special{pa 2334 380}%
\special{pa 2348 400}%
\special{pa 2334 420}%
\special{pa 2400 400}%
\special{fp}%
%
\special{pn 8}%
\special{pa 2400 200}%
\special{pa 2800 200}%
\special{pa 2800 600}%
\special{pa 2400 600}%
\special{pa 2400 200}%
\special{pa 2800 200}%
\special{fp}%
%
\special{pn 8}%
\special{pa 2800 400}%
\special{pa 3400 400}%
\special{fp}%
%
\special{pn 8}%
\special{pa 3400 400}%
\special{pa 3400 750}%
\special{fp}%
\special{sh 1}%
\special{pa 3400 750}%
\special{pa 3420 684}%
\special{pa 3400 698}%
\special{pa 3380 684}%
\special{pa 3400 750}%
\special{fp}%
%
\special{pn 8}%
\special{pa 3250 1200}%
\special{pa 3400 1200}%
\special{fp}%
%
\special{pn 8}%
\special{pa 3400 1200}%
\special{pa 3400 850}%
\special{fp}%
\special{sh 1}%
\special{pa 3400 850}%
\special{pa 3380 918}%
\special{pa 3400 904}%
\special{pa 3420 918}%
\special{pa 3400 850}%
\special{fp}%
%
\special{pn 8}%
\special{ar 3400 800 50 50  0.0000000  6.2831853}%
%
\special{pn 8}%
\special{pa 3450 800}%
\special{pa 3650 800}%
\special{fp}%
\special{sh 1}%
\special{pa 3650 800}%
\special{pa 3584 780}%
\special{pa 3598 800}%
\special{pa 3584 820}%
\special{pa 3650 800}%
\special{fp}%
\put(36.0000,-7.5000){\makebox(0,0)[lb]{$e$}}%
\put(34.5000,-7.5000){\makebox(0,0)[lb]{$+$}}%
\put(34.5000,-8.5000){\makebox(0,0)[lt]{$-$}}%
\put(34.0000,-3.5000){\makebox(0,0)[rb]{$y$}}%
\put(34.5000,-12.5000){\makebox(0,0)[rt]{$\hat{y}$}}%
%
\special{pn 8}%
\special{pa 900 798}%
\special{pa 1050 798}%
\special{fp}%
%
\special{pn 8}%
\special{pa 900 598}%
\special{pa 500 598}%
\special{pa 500 998}%
\special{pa 900 998}%
\special{pa 900 598}%
\special{pa 500 598}%
\special{fp}%
%
\special{pn 8}%
\special{pa 300 790}%
\special{pa 500 790}%
\special{fp}%
\special{sh 1}%
\special{pa 500 790}%
\special{pa 434 770}%
\special{pa 448 790}%
\special{pa 434 810}%
\special{pa 500 790}%
\special{fp}%
\put(3.0000,-7.4700){\makebox(0,0)[lb]{$w$}}%
\put(11.5000,-7.9700){\makebox(0,0){$u$}}%
\put(7.0000,-7.9700){\makebox(0,0){$F(s)$}}%
\put(20.0000,-4.0000){\makebox(0,0){$K_{\rm c}(s)$}}%
\put(26.0000,-4.0000){\makebox(0,0){$e^{-mhs}$}}%
\end{picture}%
 \caption{Multi-rate error system $\E_\mr(K)$.}
 \label{fig:multi-rate-error-system}
\end{figure}
The corresponding block diagram of the error system
$\E_\mr(K)$ is shown in Fig.~\ref{fig:multi-rate-error-system}.
Then, our problem is described as follows:
\begin{problem}
\label{prob:multi-rate-hinf}
Given target filter $K_{\rm c}(s)$, analog characteristic $F(s)$,
sampling period $h$, delay step $m$,
and upsampling ratio $L$,
find the filter coefficients $a_0,a_1,\ldots,a_{M-1}$
of $K(z)$ given by \eqref{eq:FIR} that minimizes
\begin{equation}
 \begin{split}
 \|\E_\mr(K)\|_\infty
  &=\left\|\left(\ee^{-mhs}K_{\rm c}-\hold{h/L}K(\us{L})\samp{h}\right)F\right\|_\infty\\
 \end{split} 
 \label{eq:hinf-cost-multi-rate}
\end{equation}
\end{problem}
As in the single-rate case discussed in Section 
\ref{sec:H-inf},
we use the method of fast sample/hold approximation.
Assume that $N=Lp$ for some positive integer $p$.
Then, the fast sample/hold approximation of $\E_\mr$ is given by
\begin{equation}
 E_{\mr,N}(z) = \bigl(K_{N}(z)z^{-mN} - \widetilde{H}_{N}\widetilde{K}(z)S_N\bigr)F_N(z),
 \label{eq:EmrN}
\end{equation}
where $K_{N}(z)$, $F_N(z)$ and $S_N$ are given in \eqref{eq:formula}, and
\[
  \widetilde{H}_{N} = \blkdiag\{\underbrace{\one_p,\one_p,\dots,\one_p}_L\},~
  \one_p = [\underbrace{1,\dots,1}_{p}]^\top,
\]
\begin{equation}
 \begin{split}
  \widetilde{K}(z) &= \lift\bigl(K(z),L\bigr)[1,\underbrace{0,\dots,0}_{L-1}]^\top
  =\abcd{\widetilde{A}_K}{\widetilde{B}_K}{\widetilde{C}_K}{\widetilde{D}_K},\\
  \widetilde{A}_K &:= A_K^L, \quad \widetilde{B}_K := A_K^{L-1}B_K,\\
  \widetilde{C}_K &:= \left[\begin{array}{c}
	C_K\\
	C_KA_K\\
	\vdots\\
	C_KA_K^{L-1}\\
   \end{array}\right],\quad
   \widetilde{D}_K :=\left[\begin{array}{c}
	D_K\\
	C_KB_K\\
	\vdots\\
	C_KA_K^{L-2}B_K
   \end{array}\right].
 \end{split}
 \label{eq:tKz}
\end{equation}
Note that matrices $A_K$, $B_K$, $C_K$, and $D_K$ in
\eqref{eq:tKz} are defined in \eqref{eq:FIR_abcd}.
Substituting \eqref{eq:FIR_ss} into \eqref{eq:EmrN},
we obtain
\[
 E_{\mr,N}(z) = T_1(z) + \widetilde{Q}(z) T_2(z),
\]
where $T_1(z)$ and $T_2(z)$ are given by 
\eqref{eq:T1} and \eqref{eq:T2} respectively,
and
\[
 \widetilde{Q}(z) = -\widetilde{H}_N\sum_{k=0}^{M-1}a_kz^{-k}
  = \abcd{\widetilde{A}_K}{\widetilde{B}_K}{\widetilde{H}_N\widetilde{C}_K}{\widetilde{H}_N\widetilde{D}_K}
\]
Then, using \eqref{eq:ss_pm} and \eqref{eq:ss_mult}, we have
\[
 \begin{split}
  E_{\mr,N}(z) &= 
  \left[\begin{array}{ccc|c}
 	A_1 & 0 & 0 & B_1\\
	0 & A_2 & 0 & B_2\\
	0 & \widetilde{B}_KC_2 & \widetilde{A}_K & 0\\\hline
	C_1 & \widetilde{H}_N\widetilde{D}_KC_2 & \widetilde{H}_N\widetilde{C}_K & 0
  \end{array}\right]\\
  &=:\abcd{\widetilde{A}}{\widetilde{B}}{\widetilde{C}(a)}{\widetilde{D}(a)},
  \end{split}
\]
where $a$ is the coefficient vector given by \eqref{eq:a} which is to be designed.

Finally, our problem (Problem \ref{prob:multi-rate-hinf}) is also reduced to
finding $a\in{\mathbb R}^M$ that minimizes the $H^\infty$ norm of $E_{\mr,N}(z)$,
which can be described as an LMI
from the KYP lemma (Lemma \ref{lem:KYP}) as follows.
\begin{problem}
\label{prob:multi-rate-hinf-LMI}
Find $a\in{\mathbb R}^M$ that minimizes $\gamma$ subject to
\[
 \begin{split}
  \left[\begin{array}{ccc}
   \widetilde{A}^\top X\widetilde{A} - X & \widetilde{A}^\top X\widetilde{B} & \widetilde{C}(a)^\top \\
   \widetilde{B}^\top X\widetilde{A} & \widetilde{B}^\top X\widetilde{B}-\gamma I & \widetilde{D}(a)^\top\\
   \widetilde{C}(a) & \widetilde{D}(a) & -\gamma I
   \end{array}\right]
  &<0,\\
  X&>0.
  \end{split}
\]
\end{problem}
This problem can be also efficiently solved via numerical optimization software
such as \texttt{SDPT3}, \texttt{SeDuMi}, or \texttt{cvx} on \texttt{MATLAB}.

\section{Discretization of Multi-Delay Systems}
\label{sec:multi-delay}
In this section, we consider discretization of multi-delay systems
of the following type:
\begin{equation}
 K_{\rm c}(s) = \sum_{i=1}^{\mu}e^{-m_ihs}G_i(s)
 \label{eq:Kc-multidelay}
\end{equation}
where $G_i(s)$ is a stable transfer function and
$m_i$ is a non-negative integer.
This system appears in e.g. the Smith predictor 
\[
K_{\rm c}(s)=(1-e^{-mhs})G(s),
\]
for controlling time delay systems proposed in \cite{Smi57},
whose discretization is important for implementing the controller on a digital system.
Another example is a mathematical model of multipath propagation in
wireless communications (see e.g., \cite{Gol}).

Let $K_i(z)$ be the $H^\infty$-optimal FIR digital filter
(the optimal solution to Problem \ref{prob:multi-rate-hinf-LMI})
with $K_{\rm c}(s)=e^{-m_ihs}G_i(s)$ and $m=0$.
Then the following filter
\begin{equation}
 \bar{K}(z) = \sum_{i=1}^{\mu} K_i(z)
 \label{eq:barK}
\end{equation}
is a sub-optimal FIR approximation of $K_{\rm c}(s)$ given by
\eqref{eq:Kc-multidelay} as shown in the following lemma.
\begin{lem}
Fix non-negative integer $L$ and positive number $h$.
Let $\gamma_i$ be the value of the $H^\infty$ norm of $\E_{\mr}$
defined in \eqref{eq:hinf-cost-multi-rate} with $m=0$,
$K_{\rm c}(s)=e^{-m_ihs}G_i(s)$, and $K(z)=K_i(z)$.
Then, the $H^\infty$ norm of $\E_\mr(\bar{K})$ 
with $m=0$ and $K_{\rm c}(s)$ defined in \eqref{eq:Kc-multidelay}
satisfies
\begin{equation}
 \|\E_\mr(\bar{K})\|_\infty \leq \sum_{i=1}^\mu \gamma_i.
 \label{eq:Emr_barK}
\end{equation}
\end{lem}
{\bf Proof.} First we have
\[
 \begin{split}
  \E_\mr(\bar{K}) &= (K_{\rm c}-\hold{h/L}\bar{K}(\us{L})\samp{h})F\\
  &= \biggl(\sum_{i=1}^{\mu}e^{-L_is}G_i-\hold{h/L}\sum_{i=1}^{\mu}K_i(\us{L})\samp{h}\biggr)F\\
  &= \sum_{i=1}^{\mu} (e^{-L_is}G_i-\hold{h/L}K_i(\us{L})\samp{h})F.
 \end{split}
\]
It follows that
\[
 \begin{split}
 \|\E_\mr(\bar{K})\|_\infty 
  &\leq \sum_{i=1}^{\mu} \left\|(e^{-L_is}G_i-\hold{h/L}K_i(\us{L})\samp{h})F\right\|_\infty\\
  &= \sum_{i=1}^\mu \gamma_i.
 \end{split}
\]
\hfill $\Box$

This lemma suggests if $\gamma_1,\ldots,\gamma_p$ are sufficiently small,
the sum filter in \eqref{eq:barK} is a good approximation of
multi-delay system
$K_{\rm c}(s)$ given in \eqref{eq:Kc-multidelay}
thanks to inequality \eqref{eq:Emr_barK}.

\section{Design Example}
We here show a design example of FIR digital filter design
to illustrate the effectiveness of the proposed design method.
We assume that the target analog filter is given by
\[
 \begin{split}
  K_{\rm c}(s)=&\frac{0.0031623 (s^2 + 1.33) (s^2 + 1.899)}{(s^2 + 0.3705s + 0.1681) (s^2 + 0.1596s + 0.7062)}\\
  &\quad \times \frac{ (s^2 + 10.31)}{ (s^2 + 0.03557s + 0.9805)},
 \end{split}
\]
which is a 6-th order elliptic filter with 3 dB passband peak-to-peak ripple,
50 dB stopband attenuation, and 1 (rad/sec) cut-off frequency.
This filter is computed with \texttt{MATLAB} command
\texttt{ellip(6,3,50,1,'s')}.
We set sampling period $h=1$ (sec),
upsampling ratio $L=2$ (i.e., we consider a multi-rate system),
and delay step $m=5$.
The analog characteristic $F(s)$ is chosen as
\[
 F(s) = \frac{1}{s+1}.
\]
The fast sample/hold approximation factor $N$ is chosen as $N=6$.
We fix the FIR filter length $M$ to be 32.

\begin{figure}[tb]
 \centering
 \includegraphics[width=0.97\linewidth]{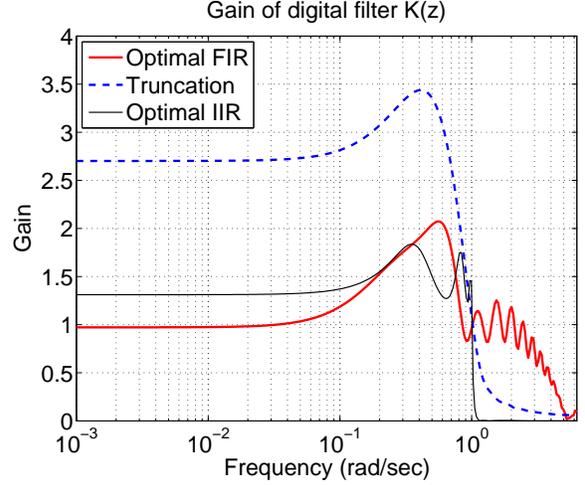}
 \caption{Digital filters: $H^\infty$-optimal FIR filter (solid thick line), 
 truncation of $H^\infty$-optimal IIR filter (dashed line), and $H^\infty$-optimal IIR filter
 (solid thin line).}
 \label{fig:filters}
\end{figure}

Under these parameters, we design the $H^\infty$-optimal FIR digital filter
based on the LMI optimization described in Problem \ref{prob:multi-rate-hinf-LMI}.
Also, we design the $H^\infty$-optimal IIR digital filter 
based on \cite{NagYam13},
and we truncate the $H^\infty$-optimal impulse response to obtain an FIR digital filter of length 32.
Fig.~\ref{fig:filters} shows the frequency response of the obtained 
digital filters.
The truncated FIR filter shows a large differences in low frequencies
while the $H^\infty$-optimal FIR filter shows differences at high frequencies.

\begin{figure}[tb]
 \centering
 \includegraphics[width=0.97\linewidth]{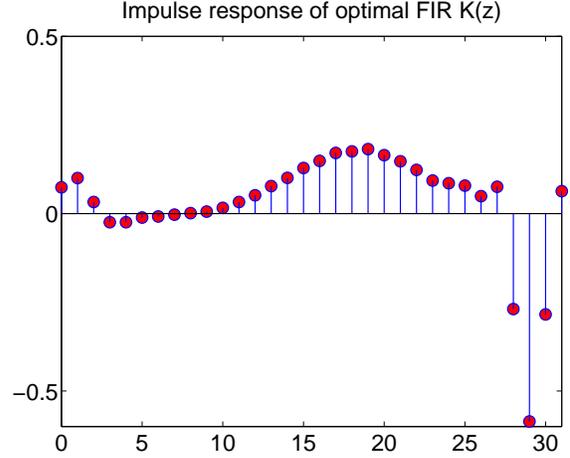}
 \caption{Impulse response of $H^\infty$-optimal FIR filter.}
 \label{fig:impulse_FIR}
\end{figure}

\begin{figure}[tb]
 \centering
 \includegraphics[width=0.97\linewidth]{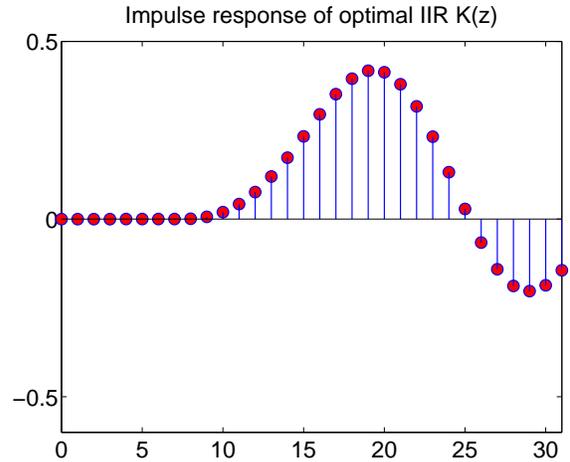}
  \caption{Impulse response of $H^\infty$-optimal IIR filter.}
 \label{fig:impulse_IIR}
\end{figure}

Figs.~\ref{fig:impulse_FIR} and \ref{fig:impulse_IIR} show
the impulse responses (filter coefficients)
of the $H^\infty$-optimal FIR filter
and the truncated FIR filter, respectively.
The $H^\infty$-optimal FIR filter has non-trivial values around $k=0$ and $k=31$,
which can not be obtained by just truncation as shown in Fig.~\ref{fig:impulse_IIR}.

\begin{figure}[tb]
 \centering
 \includegraphics[width=0.97\linewidth]{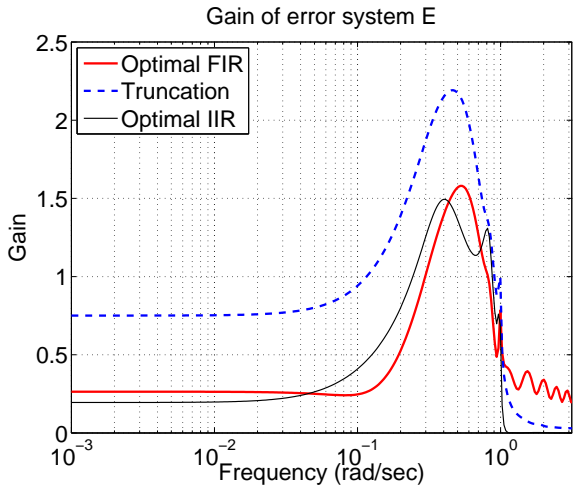}
 \caption{Gain frequency response of sampled-data error system $\E_\mr$:
 $H^\infty$-optimal FIR filter (solid thick line), 
 truncation of $H^\infty$-optimal IIR filter (dashed line), and $H^\infty$-optimal IIR filter
 (solid thin line).}
 \label{fig:error_gain}
\end{figure}

To see the difference of performance between
the $H^\infty$-optimal FIR filter and the truncated FIR filter,
we show the gain frequency response of the sampled-data
error system $\E_\mr$ defined in \eqref{eq:EmrK}
in Fig.~\ref{fig:error_gain}.
The truncated FIR filter shows a large approximation error
at low frequencies and results in a larger $H^\infty$ norm
of the error system,
while the $H^\infty$-optimal filter shows a tolerable performance
for all frequencies.
This is a merit of the use of $H^\infty$ optimization.

\section{Conclusion}
In this paper, we have proposed a method for the design
of an FIR digital filter that optimally approximates a given analog
filter with a sampled-data $H^\infty$ performance index.
The design is described as an optimization with LMI,
which can be efficiently solved by numerical optimization softwares.
We also extend the proposed method to multi-rate and multi-delay systems.
A design example has shown the effectiveness of the proposed method.
Future works include 
\begin{itemize}
\item design of FIR filters with
gain minimization on a subset of the frequency range
by the generalized KYP lemma proposed by \cite{IwaHar05}
as used in \cite{NagYam12}.
\item multiplierless implementation of $H^\infty$-optimal FIR digital filters
as discussed in \cite{Sam89,MacDem05}.
\end{itemize}



\end{document}